\documentclass[acmsmall]{acmart}

\usepackage{caption}
\usepackage{subcaption}
\usepackage{tabularx}
\usepackage{enumitem}

\begin{document}


\setcopyright{cc}
\setcctype{by}
\acmJournal{PACMHCI}
\acmYear{2025} \acmVolume{9}
\acmNumber{7}
\acmArticle{CSCW365}
\acmMonth{11}
\acmDOI{10.1145/3757546}
\received{October 2024}
\received[revised]{April 2025}
\received[accepted]{August 2025}

\title{Affordances and Design Principles of The Political Left and Right}

\author{Felix Anand Epp}
\orcid{0000-0001-6252-7244}
\affiliation{%
    \department{Department of Design}
  \institution{Aalto University}
  \city{Espoo}
  \country{Finland}
}
\affiliation{%
  \department{Social Computing Group}
  \institution{University of Helsinki}
  \city{Helsinki}
  \country{Finland}
}

\email{mail@felix.science}

\author{Jesse Haapoja}
\orcid{0000-0001-6877-7957}
\affiliation{%
    \department{Department of Computer Science}
    \institution{Aalto University}
    \city{Espoo}
    \country{Finland}
}
\affiliation{%
  \department{Social Computing Group}
  \institution{University of Helsinki}
  \city{Helsinki}
  \country{Finland}
}

\email{jesse.haapoja@helsinki.fi}

\author{Matti Nelimarkka}
\orcid{0000-0001-9867-692X}
\affiliation{%
  \department{Social Computing Group, Centre for Social Data Science}
  \institution{University of Helsinki}
  \city{Helsinki}
  \country{Finland}
}
\affiliation{%
    \department{Department of Computer Science}
  \institution{Aalto University}
  \city{Espoo}
  \country{Finland}
}
\email{matti.nelimarkka@helsinki.fi}

\renewcommand{\shortauthors}{Felix Anand Epp, Jesse Haapoja, and Matti Nelimarkka}

\begin{abstract}

Like any form of technology, social media services embed values.
To examine how societal values may be present in these systems, we focus on exploring political ideology as a value system.
We organised four co-design workshops with political representatives from five major parties in Finland to investigate what values they would incorporate into social media services.
The participants were divided into one right-leaning group, two left-leaning groups, and one mixed group.
This approach allows us to examine the differences in social media services designed by groups with different political ideologies i.e., value systems.
We analysed produced artefacts (early-stage paper mockups) to identify different features and affordances for each group and then contrasted the ideological compositions.
Our results revealed a clear distinction between groups: the right-leaning group favoured market-based visibility, while left-leaning groups rejected such design principles in favour of open profile work.
Additionally, we found tentative differences in design outcomes along the liberal--conservative dimension. 
These findings underscore the importance of acknowledging existing political value systems in the design of social computing systems.
They also highlight the need for further research to map out political ideologies in technology design.

\end{abstract}



\maketitle

\section{Introduction}


Technology design always encompasses values, a well-known fact among human-computer interaction researchers and practitioners.
To account for various values, value-sensitive design approaches were developed to help align the system's values with its user's values \citep{Friedman2009,Borning2012}.
Extensive work has been done to understand the values of children and parents in the context of parental software \citep{Mechelen2014,Nouwen2015}.
Alternatively, some design work incorporates specific values as a political statement, highlighting a social issue or envisioning alternative futures.
For example, \citet{Fox2019} focused on the distribution of menstrual products in public toilets, a feminist issue.


Researchers have often focused on end-users and less on decision-makers, such as political elites, when considering different values and their influence on design \citep[with the exception of][]{Gron2020}.
At the same time, the government is a major provider of services, facilitated increasingly through digital means.
Even more importantly, in democratic societies, the elected representatives have the legislative power -- they establish laws increasingly related to technology, as shown in the European Union's AI Act or the California Consumer Privacy Act.
Both involve \textit{value-based decisions} about the rights and responsibilities of citizens and their expected roles.
In the political system, there are reasons to suggest that such decisions mirror a wider set of beliefs and assumptions (or \textit{political ideologies}), as well as political realities based on different balances of power among representatives.
Members of Congress, Parliament, or city councils, and other representatives rarely have been studied as stakeholders in the design of digital systems. 


To this end, we organised co-design workshops with political representatives to elicit how their ideologies would be embedded in technology design.
In four co-design workshops, 13 participants across the political spectrum were tasked to design a social media platform.
We arranged participants into one right-leaning group, two left-leaning groups, and one mixed group for the workshops to compare their design and value decisions.

We examined the outcomes of the workshops as the participants' suggestions and analysed those towards the specific features and implicit affordances. Based on the literature on social media platforms and their affordances~\cite{boydSocialNetworkSites2010, Bucher2017, devitoPlatformsPeoplePerception2017, vanraemdonckTaxonomySocialNetwork2021}, we categorised affordances and connected them to underlying values to uncover ideological tensions between the different design solutions. 

We next discuss relevant literature to present political ideologies as values and discuss social media design further.
Following this, we describe the organisation of the co-design workshops.
In this paper, we focus only on the final artefacts of the co-design activity, which exemplify the value-based decisions the groups made during the design workshop.
We present a method to analyse these artefacts and summarise observations from the artefacts.
We observe that in our sample, the right-leaning group supported market-based dynamics while left-leaning groups supported diverse self-presentation.
The mixed group seemed to have both kinds of features.
Before concluding our work, we discuss the importance of ideological diversity in design activities.
Our work makes the following contributions focused on the design of information technology:
\begin{itemize}
    \item We introduce political ideologies as a set of values and connect the debate on values to national politics.
    \item We introduce a co-design approach which allows one to examine political ideologies in technology design.
    \item We show how political ideologies (right--left) correspond with a particular type of social media design and identify design principles of flexible identity and competitive logic to be specific to the left- and right-leaning parties in our sample.
\end{itemize}

\section{Background}

Our work is based on the well-established observation that technology design and development are not value-neutral activities \citep[among many others][]{Nissenbaum2005,winner85}.
\citet{winner85} infamously asked if artefacts have politics following a story about the bridges on Long Island, New York, USA. The bridges were constructed too low for public transportation buses to pass under them, thus removing access to the area for people using public transportation, primarily black people at that time.
Winner argued that this is related to the designers' social class bias and racial prejudice.
Therefore, the bridges became manifestations of their designer's political values.
In the context of human--computer interaction, \citet{Keyes2019} has similarly called out the often apolitical nature of work, which they say to be implicitly neo-liberal in nature, calling out for more diverse ideological work.

\subsection{Political Ideologies as a Set of Values}
\label{sec:political_ideologies_as_values}

Values are conceptualised in different ways, and these conceptualisations often differ between academic disciplines.
For example, political scientists often focus on political ideologies instead of values.
\textit{Political ideologies} are a set of values, beliefs, and attitudes people share that impact how they perceive and act in the world~\citep{Denzau1994}.
Instead of comprising single values, they formulate a value system.
These political ideologies have practical consequences:
within representative bodies, people with similar ideologies tend to form a political party (or a social movement in their early stages) to collaborate in advocating their values.
The existence of cohesive parties allows one to map out ideologies' dimensions; in the United States of America, this is often represented as a single dimension (republican--democrat) where various societal conflicts, such as economic stances, the role of the government, or racial assumptions, are positioned.

\begin{figure}
    \centering
    \includegraphics[width=0.33\textwidth]{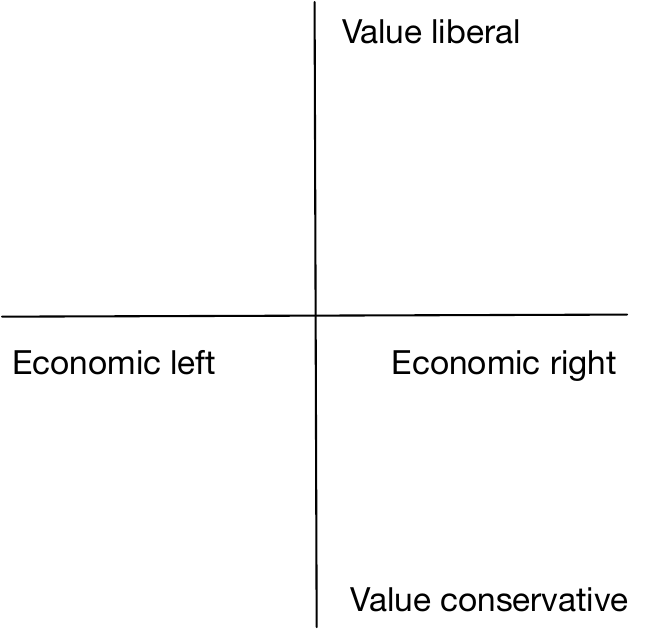}
    \caption{The two-dimensional model of political ideologies}
    \label{fig:ideology}
    \Description{Two axes crossing with one labelled value liberal and value conservative, and the other labelled economic right and economic right.}
\end{figure}

Instead of a single ideological axis, a two-dimensional model is used across Europe (where our study is situated) (see Figure~\ref{fig:ideology}).
The first axis corresponds to different values in terms of the degree of economic freedom (left--right), and the second axis corresponds to the level of desirable societal diversity (value conservative--liberal or green-alternative-liberal--traditional-authoritarian-nationalist)~\citep{Caughey2019,Kriesi2006}.
For example, a supporter or member of a left-liberal party, such as social democrats or the greens, might push for the right to self-define gender identity and extensive government oversight in the economic operation.
In contrast, a supporter or member of a right-conservative party, such as Christian democrats or conservative parties, might push for traditional gender roles and limit the government's role in the economic affairs.
In general, while the model is simplistic, it has provided extensive value for empirical research on political decision-making, political opinions, and voting \citep[e.g.,][]{Caughey2019,Huber1995,Hooghe2002,knight2006fix}.
For example, political scientists have identified what kind of social and economic policies are commonly used across the two-dimensional model \citep{COCHRANE2019,Jahn2011,Daubler2022}.
In addition, some research has focused on the interconnection between technology and the party divisions:
There is mixed evidence that left-leaning political parties are adapting information and communication technology more extensively to allow contact and participation in party functions \citep{Gonzalez-Cacheda2024}.

As ideologies are systems based on a set of values, these values can manifest themselves in the design and use of digital services.
Examining the design of social media services, \citet{Gron2020} illustrate that the left--right axis relates to different degrees of paternalistic values: left-leaning participants supported a higher degree of paternalism while right-leaning participants highlighted the importance of individual responsibility.
Their work highlights ideologies as an area of political competition, where individuals with different ideologies prefer design approaches aligned with their worldview. 
Relatedly, \citet{Detienne2019} built on the concept of ideologies to understand community memberships in conflict.
For example, they identified an ideological conflict in the editing of Wikipedia articles.
They observed strong disagreement between authors with a religious point of view and those with a scientific point of view, leading to an edit war and personal attacks on the platform. This instance underscores the potential for conflict and disagreement between political ideologies, making it clear that finding a compromise on an agreed solution can be difficult and sometimes impossible.
This work highlights the importance of reflecting more expansive sets of values (or ideologies) and understanding them through wider groups, even if their ideologies do not fit the two-dimensional political ideology model presented above.

The above examples illustrate the value of approaching not only individual values but sets of values, in other words, ideologies, in contrast to more prevalent studies focusing on specific values in context of information system design.
The focus on political ideologies -- pushed forward by political parties -- provides the additional benefit of reflecting the legislative and executive processes at the societal level.

\subsection{Values in the Context of Social Media Services}

Like any sociotechnical arrangement, social media services incorporate and display values.
The most prominent value tension in contemporary public discourse concerns social media companies' for-profit operations and their contrast with other values, such as human rights.
Thus, the discourse has a focus on capitalism.
For example, \citet{zuboff} has argued that we are now in the era of surveillance capitalism, in which the collection and use of personal data has become a widespread business model.
Social media companies use such personal data for targeted advertising, but while doing so, limit our right to privacy.
An alternative line of criticism is the focus on user engagement and attention as a (capitalist) metric, leading to the promotion of affective content and thus having negative implications for political discourse on these platforms.

Given these concerns, there are proposals on alternative value bases for social media services.
\citet{Jia2024} discuss how the news-feed algorithm could promote content more aligned with democratic ideals by sorting content so that it does not show support for undemocratic practices, partisan animosity or violence, or social distrust.
Others have proposed different normative bases for rethinking social media services \cite[for summary, see][]{Goldberg2016}.
For example, the ideas that social media services could serve as a place for political deliberation have been repeated in the literature \citep[e.g.,][]{Semaan2014,Semaan2015}.
Similarly, researchers have shown that these values are culturally situated; what might be accepted in one cultural context may be rejected in another \citep[e.g.,][]{Nelimarkka2019} and other researchers have called for more participatory and democratic approaches to governance \citep{Fan2020,Zhang2020}.
These examples again illustrate the critical role of values as a standing point for social media service design.

However, beyond the above-mentioned \citet{Gron2020}, social media design research has not examined values through the lens of political ideologies.
Instead, it seems that political ideologies form a design challenge in the context of social media service design: how to ensure that people across the ideological spectrum interact to create common ground \citep[e.g.,][]{Combs2023,Feltwell2020,Garimella2017,Nelimarkka2018a}.
Alternatively, social media research sometimes positions ideology as a background variable that explains human behaviour or the content produced \citep[for example,][]{Morgan2013,Zhang2015a}.

Our brief review above highlights the role of values in the debate on social media service design.
It critiques the current situation and suggests alternative design approaches.
At the same time, we highlight that, from this perspective and more widely in human--computer interaction research, these values are rarely understood in terms of political ideologies.

\section{Approach, Data and Methods}

Our design activity focused on social media services as these are increasingly critical for information dissemination and political discourse in our society.
Our form of co-design aligns with ``co-creative design'', where the collaborative creativity of participants and designers creates value, which, in our case, are insights towards the research inquiry.
Here, participants are not automatically stakeholders~\citep{zamenopoulosCodesignCollaborativeResearch2018,prahaladCocreationExperiencesNext2004}. 

\subsection{Research Setting}

This study is situated in Finland, a Nordic country with a multiparty political system with 17 registered political parties, of which nine have representatives in the Finnish Parliament.
Like in the rest of Europe, the media and political scientists use the two-dimensional ideological model (Figure~\ref{fig:ideology}) to categorise these parties \citep{gronlundvalue}.
In recent years, the ideological dimensions have become more prominent and affective polarisation has increased \citep{Kekkonen_Kawecki_Himmelroos_2024}.

In terms of political power, there are three equally large parties, each with about 20\% of the popular vote in the most recent elections: 
National Coalition Party,
the Finns Party,
and Social Democratic Party.
Following \citet{gronlundvalue}, National Coalition Party and the Finns Party are political right, whereas the Social Democratic Party is economic left.
The Finns party is the only value-conservative party in this group, while both National Coalition Party and the Social Democratic party belong to the value-liberal side of parties.
Following them, there are three somewhat smaller parties, with between 5\% and 10\% of the popular vote:
the Center Party,
the Greens Party,
and the Left Alliance,
The Greens party and the Left Alliance are left-leaning and value-liberal parties,
while the Centre party is right-leaning and the only value-conservative party in this set of parties.
All parties above have elected municipal, national and European parliament representatives and have been part of the current or previous national government.
The three remaining parties
Swedish People's Party of Finland,
Christian Democrats,
and Movement Now, all had less than five per cent of the popular vote.
Ideologically, the first one belongs to the economic right and value-liberal quadrant, while the second belongs to economic right and value-conservative quadrant;
the sample is too small to make definitive position for the Movement now, but within popular discussion, it is a value-liberal economic right party.

When positioned on the two-dimensional model of political ideologies (Figure~\ref{fig:ideology}), the parties are situated in the
value-liberal economic left,
value-liberal economic right,
and value-conservative economic right quadrants.
Beyond these, there are various specific issues for each party; for example, the Swedish People's Party of Finland strongly supports Finland as bilingual country (with both Finnish and Swedish serving as official languages).
While these specific issues are not captured by the two-dimensional model, the two-dimensional model is often used to describe the ideological landscape in Finland \citep[e.g.,][]{gronlundvalue,Kekkonen_Kawecki_Himmelroos_2024,Ford2020,Huber1995}.

\subsection{Participants}

In total, we organised four workshops with 13 participants representing five\footnote{ We reached out to the six largest parties, but even after several attempts, one party did not respond to our request.
This also limited our ability to organise mixed and right-leaning workshops.\label{footnote:parties_not_workign_with_us}} of the six major parties.
These parties included both right-leaning and left-leaning parties and included parties with more centralist stances and more extreme stances.
These parties represent the Finnish political spectrum broadly: together these five parties gained more than 60\% of the popular vote, with the missing party adding an additional 20\% of popular vote to the mix; thus our original target represented 86\% of total popular vote.
We contacted the party's general secretary, executive director, or similar high-level officials.
We asked them to suggest candidates for the workshop with expertise in social media and familiarity with the party's political positions.
Their suggestions included city council members, party office staff and youth organisation members, who we would consider to have a clear political ideology (we further confirmed this with a survey, as indicated later).
Of the participants, six identified as women, six identified as men, and one identified in the other gender category.
Their ages ranged from the early twenties to the seventies, with a median of 34 and a mean of 39.
As Finland is a relatively small country, we do not provide more detailed demographic details to ensure the anonymity of participants.

\begin{figure}
    \centering
    \includegraphics[width=0.5\linewidth]{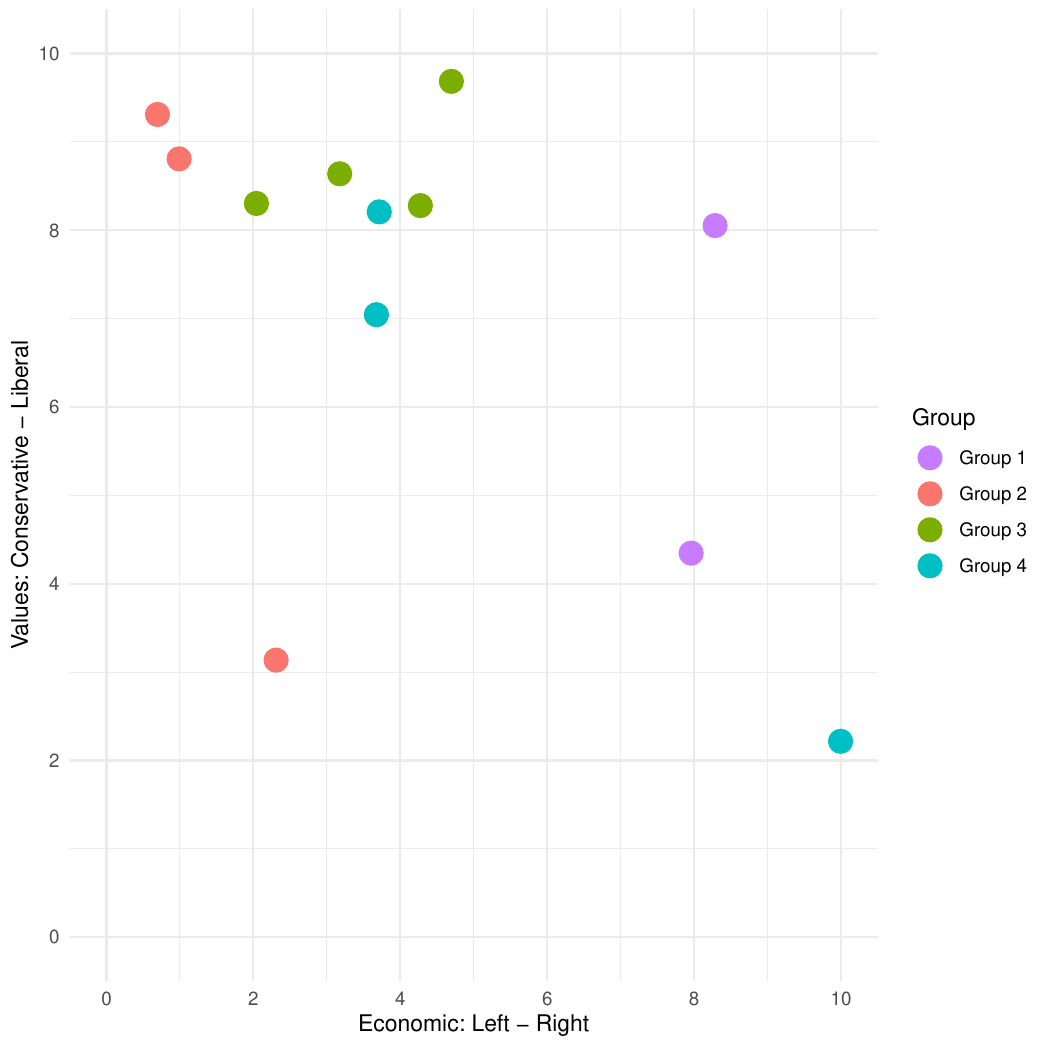}
    \caption{Position of workshop participant on the ideological map.}
    \label{fig:participant_positions}
    \small{Positions adjusted with jitter to avoid overlaps.}
    \Description{A chart with the coloured dots across the two-dimensions value liberal--conservative and economic right--left.}
\end{figure}

The candidates suggested by party officials were suitable participants for this study.
First, they have experience discussing politics through computer-mediated means and face-to-face settings.
All participants discussed politics daily or often in face-to-face settings.
The majority also discussed politics daily or often through instant messaging apps (5 daily, 6 often and 2 sometimes) and on social media (7 daily, 2 often and 4 sometimes).
Second, they had an established ideological stance.
All participants had a well-established political party identity, i.e. there was a party they felt closest to.
In addition, we conducted a survey of political ideologies on the economic left---right and value conservative--liberal axis\footnote{ We used the survey items used in the most recent Finnish National Election Survey to evaluate ideological positions and discussed with the project lead of the National Election Survey on the suitability of these indicators.} to confirm that these individuals self-described their political ideology similar to the ideology position of their party (see Figure~\ref{fig:participant_positions}).

\begin{table}
        \centering
    \begin{tabular}{lcccc}
    \toprule
         & Group 1 & Group 2 & Group 3 & Group 4 \\
         & Right-leaning & Left-leaning & Left-leaning & Mixed \\
    \midrule
     Economic left (0)-- Economic right (10) & 8.0 & 1.33 & 3.50 & 5.50 \\
                 & (0.00) & (0.577) & (1.29) & (3.00) \\
    Value conservative (0)-- Value liberal (10) & 6.00 & 7.00 & 8.75 & 6.75 \\
                 & (2.83) & (3.46) & (0.96) & (3.40) \\
    \bottomrule           
    \end{tabular}
    \caption{Mean and standard deviation of ideological measurements per group.}
    \label{tab:group_compositions}
\end{table}

The groups had varied self-reported ideological compositions (see Table~\ref{tab:group_compositions}):
one consisted of participants solely right-leaning (Group 1),
two consisted of participants solely left-leaning (Groups 2 and 3),
and one was mixed, including participants both right- and left-leaning (Group 4).
With the exception of Group 1\footnote{Unfortunately, this group had a no-show from the other right-leaning party in our participant pool. This led to the unfortunate situation that this group was homogeneous in terms of political parties.
As the logistics of the workshops were complicated, we nevertheless opted to run this group to ensure that we have at least one right-leaning group in our sample.}, all groups had participants from several parties.
This group composition allowed us to examine the elicitation of political preferences in the left--right leaning axis, thus engaging how ideological thinking impacts research activities.

\subsection{Organisation of Co-design Workshops}

The co-design workshop lasted about 2--3 hours and consisted of five activities altogether (see Table~\ref{tab:schedule}).
The first two activities were individual orientation and thinking activities.
First, participants were asked to attune towards the problems of social media and after they were asked to experiment with the materiality of the social media service by considering its issues and challenges primarily through a user interface, tasking users to consider its modifications.
These activities were designed as individual ones to reduce group--think and peer pressure, allowing participants to form their own perspectives.
Following the two individual activities, the participants were asked to share their thoughts and observations with the group and briefly introduce themselves if they had not done so already.
This activity allowed all participants to share perspectives and observations without interruption.
Following this, the participants moved on to the co-design activity, where they were assigned to design a new social media service in about 45--60 minutes.
We outline the problem statement around political polarisation and decreasing political discussion culture, with the following brief:

\begin{table}[b]
    \centering
    \begin{tabular}{p{.2\textwidth}p{.7\textwidth}r}
        \toprule
        Activity & Assignment & Minutes \\
        \midrule
        Individual orientation activity &
        The political debate on social media presents a range of opportunities and challenges. What would you highlight as challenges and opportunities? Participants were instructed to use sticky notes during, with one idea per sticky notes. & 10 \\
        Individual design thinking activity &
        What Twitter as a service could do to reduce the challenges and highlight the better opportunities. Think about this through the different perspectives of Twitter: what should change? Participants were given mock-up illustrations from Twitter and instructed to make notes and draws over them. & 10--15 \\
        Sharing activity & Please share key observations and highlights to the group. & 15 \\
        Co-design activity & Design a new social media service to address issues on social media you have observed. (More extensive brief presented inline.) & 45--60 \\
        Presentation & Please present the prototype to the research team and answer their questions. & 15--25 \\
        \bottomrule
    \end{tabular}
    \caption{Overview of the co-design workshop schedule}
    \label{tab:schedule}
\end{table}

\begin{quote}
Your task is to work together to design a new service for people who want to discuss politics: PoliPulse.
The service is aimed at politically active Finns to exchange opinions and reduce polarisation in society.
The aim is to avoid the harmful features of other social media and to improve the functionality and atmosphere of social debate.
You can only post text and links to the service.
Your task as a group is to create an initial prototype of this service, showing the main interfaces and being able to explain what users do in these interfaces.
\end{quote}


During the workshop, we did not want to intervene with the group dynamics and reasoning processes, so the research team did not facilitate the co-design activity beyond providing assignments and time management.
The exception was ensuring that all participants had their say during the sharing activity if the groups were not doing this themselves.
During the co-design activity, we provided the groups with pre-made icon sets and neutral figures to reduce the need for graphic design skills; they were also given empty user interface canvases and pens.
We did not elaborate on their interpretations of the assignments or design briefs beyond practical considerations -- such as reminding people to use one sticky note per idea, use legible handwriting, or introduce the design materials.

In the final step, each group needed to present their joint proposal, which demanded that groups consolidate which ideas to include in their final artefact.
We were primarily interested in these artefacts and their presentation as the basis for our analysis, as they should exemplify ideological tendencies in the outcomes of a design process.

In addition to the co-design activity, participants filled in a pre-survey (collecting basic demographics data) and an exit survey, asking about their experiences of the co-design workshop.

\subsection{Ethical Considerations}

Our methods borrow from the long-standing traditions of participatory design, so we want to reflect on our use of collaborative design as a research method. 
In our case, we purposely gathered members of the political elite instead of any under-represented group, aligning with the tradition of `studying up'.
As \citet{Nader1972} has highlighted, focusing research on those in positions of power opens up new ways to theorise and informs democratic societies on the role of the powerful.
As we introduce novelty in using co-design workshops to study connections between politics and design, we caution other researchers to use these with respect to the political realities in place. 

\subsection{Analysis}

The primary aim of this work was to highlight how political ideologies become embedded in the design process.
The workshop we conducted served to gather the cultural norms, assumptions, and ideologies of the co-designers, here the political representatives, by capturing a design process.
Co-design activities include examining various design directions and ideas; however, for this work, we wanted to focus on each group's converged presentation.
Therefore, we focus only on (1) the finished co-designed artefacts, i.e. sketches on potential user interfaces (see Figure~\ref{fig:sketches} for an example) and (2) groups' presentations on the artefacts and their operations in this analysis; i.e. the final step of the co-design workshops.
For analysis purposes, the artefacts were scanned and final presentations were transcribed verbatim.

\paragraph{Artefact analysis}

We used qualitative content analysis to identify affordances from these sketches~\citep{rodgersUsingConceptSketches2000}.
We build on the Feature Analysis method~\citep{Hasinoff2021} to connect this analysis to ideologies to examine design artefacts for their value-loaded contents.
\citet{Hasinoff2021} developed the Feature Analysis as a qualitative content analysis that identifies ideologies in mobile applications based on their features.
Their work builds on an inductive approach to identifying features and later categorising them into affordances.

Naturally, the question becomes what a feature or an affordance are and how we interpret them.
For \citet{Hasinoff2021}, a feature was ``a function that users control or are likely aware of''.
In contrast, for them an affordance was a higher level conceptualisation of a characteristic of the technology that ``request, demand, allow, encourage, discourage, and refuse"~\cite[p. 241]{davisTheorizingAffordancesRequest2016} a user behaviour.\footnote{~This list of verbs is an extension from the most classical notations of an affordance in human--computer interaction, where only two verbs might be used.
We acknowledge that affordances have been widely used in human--computer interaction and computer-mediated communication research \citep{Evans2017,Hutchby2001}, with various definitions.
\citet{Hasinoff2021} builds on \citet{{davisTheorizingAffordancesRequest2016}} recent work on affordances, which has focused on a wider set of actions.
}


As such, an affordance describes the relationship between the system and the user on a more abstract level.
As they promote or sanction specific user behaviours, they describe the norms set by the design. 
To examine affordances and their implicit norms specific to social media services,
we examined works which classified affordances \citep{boydSocialNetworkSites2010,Bucher2017,devitoPlatformsPeoplePerception2017,vanraemdonckTaxonomySocialNetwork2021} to sensitise our analytical work on pre-existing abstracted affordances.
These have included for example affordances related to
self-presentation~\cite{devitoPlatformsPeoplePerception2017},
communicative functions~\cite{Bucher2017},
and persistence~\cite{boydSocialNetworkSites2010}.
Two authors identified individual features through open coding of the material in multiple iterations of going over the material and discussing and refining codes following common qualitative coding practices.
We then categorised those features into abstract affordances based on the existing literature. 
By examining the features and examining suitable higher-level abstractions, we ended up focusing on the affordances \textit{visibility}, 
\textit{network interactions} and \textit{intervention} \cite{vanraemdonckTaxonomySocialNetwork2021}.
In addition, we analysed which affordances structured the user's self-presentation based on the framework by~\citet{devitoPlatformsPeoplePerception2017} under the category of \textit{profile work}, a concept we adopted from \citep{silfverberg_ill_2011, uskiSocialNormsSelfpresentation2016} which refers to self-presentational labour done in social media settings in contrast to face-to-face situations.

\paragraph{Analysis of the final presentations}

In addition to artefacts, we used a similar qualitative analysis process to examine the video recordings and verbatim transliterations of the presentation activity.
This was the final step of the co-design workshop, where participants were asked to present the function of the service (using the artefacts) and the research team asked clarifying questions.
We used this material to examine how groups motivated particular features and aligned our interpretation of the addition of said features (and related affordances) to how they described its intended use or reasoned its addition.

\paragraph{Mapping affordances and ideologies}

\begin{figure}[b]
    \centering
    \includegraphics[width=1\linewidth]{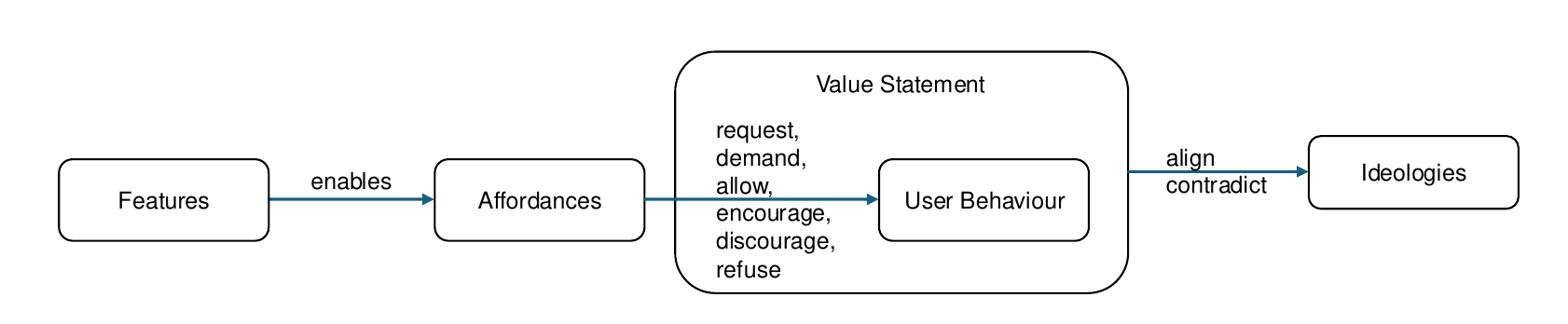}
    \caption{Our analytical lens identified affordances and their implicit values to compare those to ideologies.}
    \label{fig:analytical_framework}
    \Description{A diagram with text boxes and arrows pointing from left to right: Features enable affordances, affordances request, demand, allow, encourage, discourage, and refuse user behaviour which is encapsulated in a box stating value statement. This points towards to align or contradicting Ideologies.}
\end{figure}

Using our understanding of how particular affordances describe normative relations to user behaviour, we consider them as consolidated value statements of each group (see Figure \ref{fig:analytical_framework}).
These value statements then align or contradict the behaviour in accordance to a political ideology.

Our final step was to map affordances into political ideologies.
There is no established practices for this:
\citet{Hasinoff2021} argues their research approach seeks to analyse ideologies, their analytical move from low-features and higher abstraction affordances (as described above) to ideologies is not clearly described.
They suggest using speculative design approaches to map out alternatives from different ideological stances, however it remains unclear how ideologies are integrated into this analysis
(for further highlights on this Section~\ref{sec:method_discussion}.)

Instead of seeking to interpret ideologies from the artifacts and material  we contrasted the affordances across groups based on participants'  \textit{self-identified} political ideology.
In addition, we used \textit{existing literature} from political science on the economic left--right and value conservative--liberal dimensions\footnote{~While we acknowledge there are various ways to define political ideologies (see Section~\ref{sec:political_ideologies_as_values}), for this analytical stage we focused on these axes to clearly define what is meant by ideology in this work.
Our interpretations on ideologies builld on previous work on these axis in Finland and Europe, such as those by \citet{COCHRANE2019}, \citet{Jahn2023}, \citet{Daubler2022} and \citet{gronlundvalue}.} to establish the connection between choices on affordances and ideologies; or seek a connection with the identified affordance and political ideologies..
Naturally, there are other factors beyond party alignment (discussed in more detail in \ref{sec:method_discussion} and \ref{sec:limitations}), this evidence provides justified interpretations of the role of political ideologies in the design of services; showing how we see a connection between an affordance, its potential description by participants, and political science work on ideologies.
Therefore, as the final step we compared affordances between groups to identify patterns on affordances following political ideologies and connect them to previous work on ideologies.

\section{Results}

Our analysis identified differences
in the overall system architecture,
functionality around posting and showing content,
and
profiles.
These sections are intentionally kept `close to the data' and avoid drawing interpretations to allow us to demonstrate the case.
The final subsection abstracts the design choices towards their affordances and brings ideological differences into the analysis.

\subsection{General Architecture}

\begin{figure}
    \begin{subfigure}{0.5\textwidth}
    \centering
    \includegraphics[width=\textwidth]{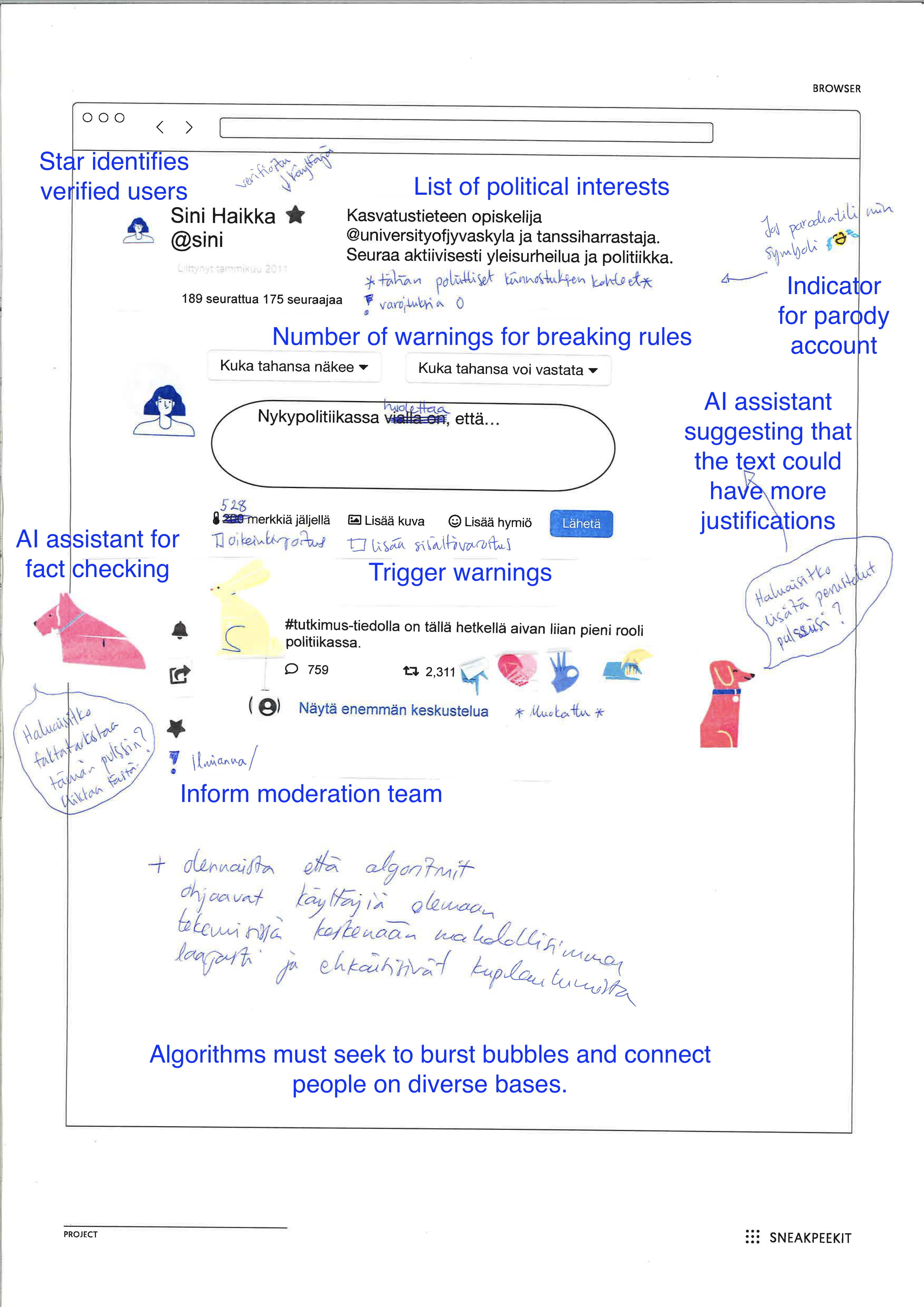}
    \caption{Group 3 news feed}
    \Description{A sketch showing a news feed designed by group 3}
    \label{fig:result-group3}
    \end{subfigure}~\begin{subfigure}{0.5\textwidth}
    \centering
    \includegraphics[width=\textwidth]{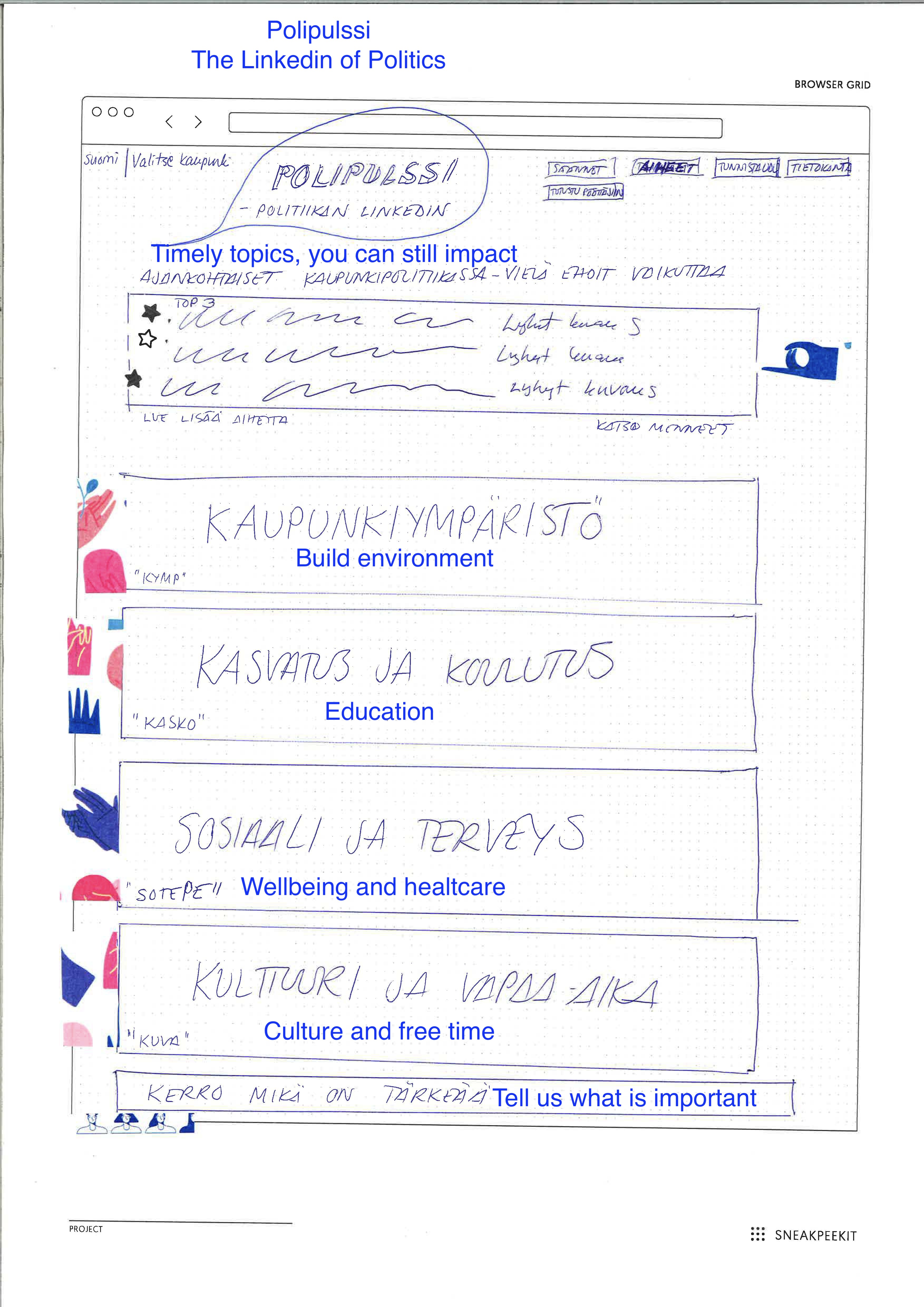}
    \caption{Group 1 forum main page}
    \Description{A sketch showing a forum designed by group 3}
    \label{fig:result-group1}
    \end{subfigure}

    \caption{Main pages of two different groups}
    \label{fig:sketches}
\end{figure}

The design brief allowed for different interpretations of the goal.
Groups 1, 2, and 4 presented artefacts more like a traditional Internet forum, while Group 3's artefact mimicked social media with a news feed.

Within forums, architectures varied.
Group 1 structured the discussion forum through organisational units (such as health and welfare services, organised publicly and locally in Finland), followed by sub-forums based on ongoing projects (see Figure~\ref{fig:result-group1}).
Group 2, on the other hand, structured the discussions into three groups:
political parties,
wider civic society,
and other discussion.
Similarly, Group 4 had a mixed approach
based on thematic and location-based structures.
With the structuring of the forum, Groups 1 and 4 added elements allowing citizen initiatives on establishing new sub-forums, showing a somewhat more citizen-centric approach to participation, going beyond just administratively defined forums.
Group 3's news feed-based design (see Figure~\ref{fig:result-group3}) did not have a preset structure but used hashtag-based taxonomy to allow user-generated structuring of the content.

\subsection{Content and Posting}

\begin{figure}
    \centering
    \includegraphics[width=0.5\linewidth]{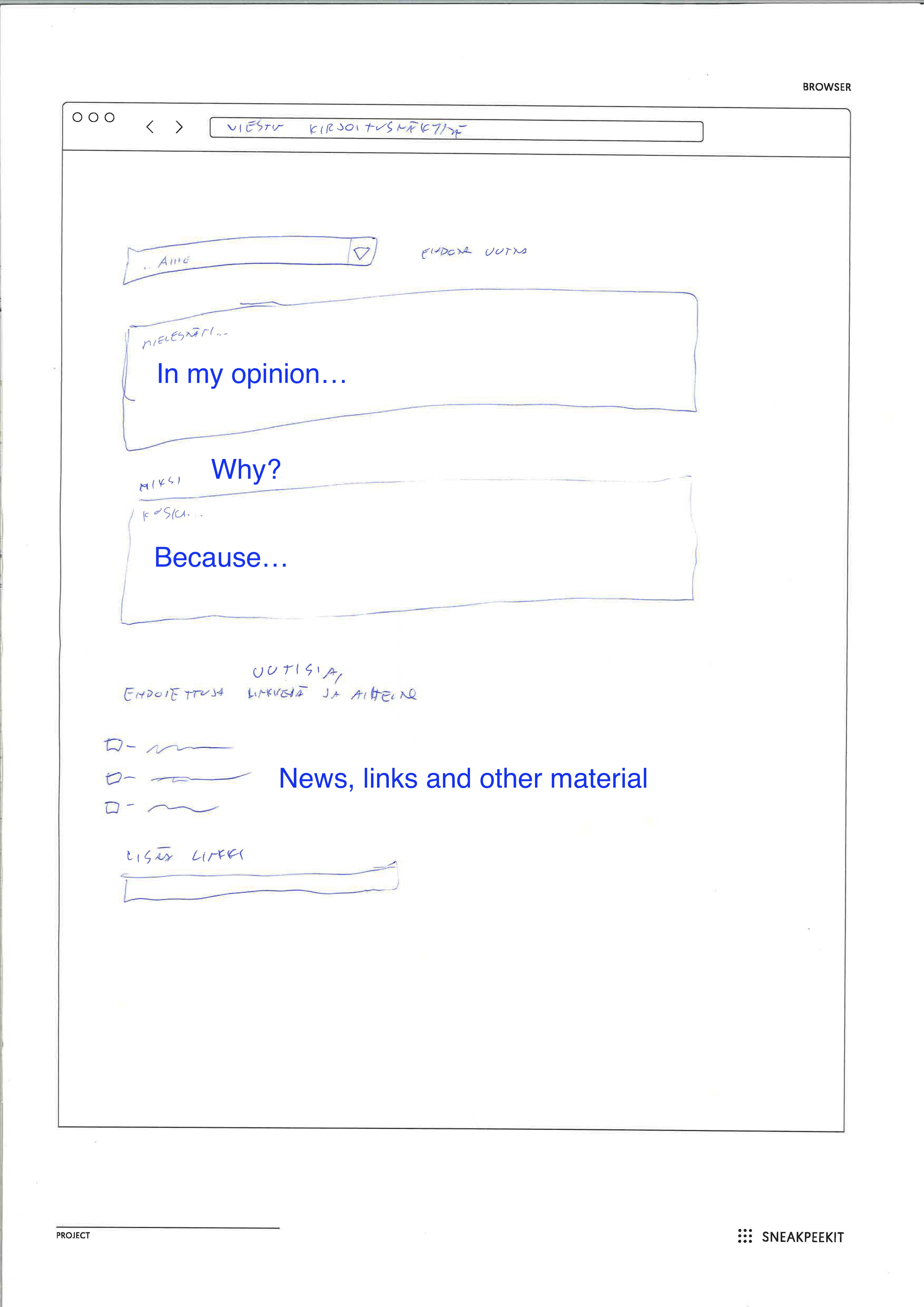}
    \caption{Scaffolding posts}
    \label{fig:scaffolded_posts}
    \Description{A sketch showing a content posting page with separated boxes for presenting ones' opinion, reasoning and supportive material.}
\end{figure}

Only groups 3 and 4 specified how new content would be submitted to these forums.
Both focused on more structured input opportunities to help citizens justify their positions.
Group 4 designed an interface which scaffolds the post generation by asking the user to enter their opinion and justifications on separated text input, with the assumption that this would lead to a higher quality debate (see Figure~\ref{fig:scaffolded_posts}).
In addition, they asked people to provide links to support their stance and recommended news content related to the opinion.
Group 3 identified the same issue but considered an artificial intelligence agent (shown as a dog in Figure~\ref{fig:result-group3}) to suggest to the user that they could provide justifications. Their system also provided users with controls for who could view or react to their content.
Interestingly, Group 3 introduced a character limit, focusing on shorter messages and propagation.


Beyond providing written content and comments, two of the artefacts included a mechanism to react to the content.
Group 1's artefact used stars to show support, and such reactions would increase the topic's visibility on lists, stating that users would be instructed to:

\begin{quote}

Presenting participant: [Describing user guidelines for researchers] Like topics which you find important so that they rise to the discussion [further explaining the sentence to researchers] so it uses the \textit{Reddit approach} [emphasis ours]

\end{quote}

As they explained, there was competition for visibility similar to the Reddit platform; where a highly upvoted content gains more visibility.
Group 3 provided various reaction icons and the ability to recirculate content on their artefacts, which they explained to be used to indicate some kind of social support towards the posts' author:

\begin{quote}

Researcher: Do I see correctly, that you have some kind of like buttons there?

Presenting participant: Yes

Researcher: Why did you want to add the like buttons?

Presenting participant: We wanted to ensure this [platform] had a positive atmosphere (laugher) and we wanted to make it so that you could \textit{show your support to others using these reactions} [emphasis ours]


\end{quote}

This articulation is arguably different from Group 1's, as Group 3 did not consider reactions a mechanism for gaining popularity. Rather, they emphasised their meaning for socio-emotional support in the final presentation.
In fact, they chose the reactions in a manner that they should not be able to be appropriated for displaying negative attitudes. 
Groups 2 and 4 did not introduce this kind of feature to the artefacts,
with Group 2 articulating how reactions, such as `likes` are not beneficial for constructive discussion and lead to attention seeking, explaining in the final presentation that:

\begin{quote}

Presenting participant:
But it’s precisely the idea that people questioned the significance of having someone who gets like 3,000 likes and that someone gets no likes, that people would focus on the discussion and not how it's said, dopamine spike.


\end{quote}



Recent academic discussions on social media have focused on content moderation and its challenges ~\citep{rantasilaOutliningApproachesImprove2022}. 
Oonly groups 1 and 3 reported any form of content moderation in the final artefact, even though the focus was to devise an online space for political discussion.
Group 1 hints at rules for the platform, which are enforced by moderators selected by some authority, which exemplifies the intention to govern the content~\cite{rantasilaOutliningApproachesImprove2022}. 
While it remained unclear who is part of the moderation team, the moderation must be conducted as a team effort.
Group 3 described a more user-driven approach to moderation with options to report posts, add trigger warnings, and show reminders to enforce rules and block users.
Similarly, Group 3 mentioned that there is a human editor to monitor automated fact-checking. This shows a mix of hard (e.g., removal of reported posts) and soft (e.g. trigger warnings) moderation styles~\cite{rantasilaOutliningApproachesImprove2022}.
While we can surmise rules against spam and hate speech from Group 3 suggesting these as features, no group clearly described what content is warranted or sanctioned on their sites.

\subsection{Profiles}

\begin{table}[b]
\begin{tabular}{lcccc}
\toprule
                               & Group 1 & Group 2 & Group 3 & Group 4 \\
                               \midrule
Picture                        & \checkmark & \checkmark       & \checkmark       & \checkmark       \\
Verification of authentication & \checkmark & \checkmark & \checkmark       &    \checkmark      \\
Username                       &         & \checkmark       &         &         \\
Real name                      & \checkmark  &         & \checkmark       & \checkmark       \\

Description                    &         &         & \checkmark       & \checkmark       \\
Age                            &         &         &         & \checkmark       \\
Party affiliation              &         & \checkmark       &         &         \\
Occupation                     &         &         &         & \checkmark       \\
List of political interests    &         &         & \checkmark       & \checkmark       \\
Follower counts                &         &         & \checkmark       &         \\
Indicator for a parody account &         &         & \checkmark       &         \\
Number of warnings             &         &         & \checkmark       &  \\ 
\bottomrule
\end{tabular}
\caption{Design of user profile across groups}
\label{tab:user_profile}
\end{table}

All four designs contained the idea of a persistent user identity as a core part of the platform.
At the same time, there was a lot of diversity in the details that should be included in the artefacts.
All sites included some profile pictures.
Group 2 allowed pseudonyms and a person to change their username, which would keep the posting history intact.
Instead, groups 1, 3 and 4 enforced a real-name policy.
However, Group 3 hinted at the possibility of parody accounts, although marked by an indicator.
The policy of enforcing clear names is often debated also on social media services such as Facebook \citep{haimsonConstructingEnforcingAuthentic2016}.

User verification was discussed in all final presentations.
Group 1 argued that the system should have strong authentication during the final presentation to support the use of the data and build up a stronger accountability for participants' actions:

\begin{quote}

Researcher: Can I ask at this point why you ended up with strong authentication?

Presenting participant: Because we have a social media platform that is used as an element in official decision-making.
[- -]

Presenting participant: Of course, to the extent to which strong authentication increases this status in the minds of many.
It ensures that it cannot be a random caller to a gossip magazine, or a bully or so on. [The authentication] controls for this.
This is where we ended up in this case.


\end{quote}
 but it also was an approach to ensure proper behaviour on the platform, and Group 3 argued in a corresponding manner that there should be a mechanism to ensure the poster `is a real person' and not pretend to be some other person.
Similar arguments were also presented by Group 2 and Group 4.

Interestingly, the differences do not stop here (see Table~\ref{tab:user_profile}).
There was variance in what aspects of users' identities were made relevant in the profiles, spanning from elements such as free self-descriptions to pre-given political affiliations and even to proofs of deviating from platform's rules.
Profiles included a free-form description field in groups 3 and 4, allowing for freedom on identity statements.
Group 2 did not design such an element; instead, after the user name, an indication of the user's party affiliation was shown.
Group 4 added age and occupation in addition to the free-formed field, as well as a list of users' political interests.
Group 3 instead included, for example, the number of warnings the user had received for breaking any rules.

\subsection{From Features to Affordances to Ideologies}
\label{sec:results-affordances}

The following section describes the differences and similarities of the artefacts along four key affordances (each row in Table~\ref{tab:affordances}).
In the following paragraphs, we will go through each key affordance and the differences and commonalities between groups, along with their ideological leanings. 

\begin{table}[b]
\begin{tabular}{lllll}
\toprule
Affordances &
  Group 1 &
  Group 2 &
  Group 3 &
  Group 4 \\ 
  & Right-leaning & Left-leaning & Left-leaning & Mixed \\
  \midrule
Popularity-based Visibility &
  Encourage & 
  Refuse & 
  Discourage & 
  Not addressed 
  \\ 
Open Network Interactions &
  Refuse & 
  Refuse & 
  Encourage & 
  Refuse 
  \\
User Intervention &
  Allow & 
  Not addressed &
  Encourage & 
  Encourage 
  \\ 
Flexible Profile Work &
  Discourage & 
  Allow & 
  Encourage & 
  Allow  
  \\
 \bottomrule
\end{tabular}
\caption{Key Affordances of the design artefacts of each group (including their ideological composition).}
\small{According to \citet{davisTheorizingAffordancesRequest2016}: Encourage prompts specific actions while suppressing others;
Allow remains neutral, indifferent to the action's usage and outcome;
Discourage makes an action more difficult to achieve; 
Refuse blocks certain actions outright.}
\label{tab:affordances}
\end{table}

The groups varied in their stance towards the mechanics of \textbf{content visibility} on their platforms:
the right-leaning 
group preferred popularity-based metrics to promote content on the platform.
Posts would have a like count, prioritising their visibility on the platform and thus introducing a competitive logic for attention: each post would compete with others in gaining reactions.
Therefore, the group \textit{encouraged} popularity-based visibility by fostering a usage pattern that aims for posts that receive more reactions and consequently get rewarded with more visibility, similar to platforms like Reddit and Facebook~\cite{vanraemdonckTaxonomySocialNetwork2021}.
Both left-leaning Groups 2 and 3 \textit{rejected} popularity-based design approaches.
As we observed above, while Group 3 mentions the need for reactions and likes, they highlight them as easy ways to show support, not as a popularity-based logic.
Therefore, Group 3 \textit{discouraged} popularity-based visibility, but still had some affordances which may make it possible, whereas Group 2 outright \textit{refused} it by not offering any affordance for users to engage with content based on its popularity.

Regarding \textbf{interaction} forms, all but Group 3 opted for a subforum architecture. 
Group 1, 2 and 4 all decided to section the discussion on their platform according to organisational levels (Group 1);
parties, society and others (Group 2); and themes or location (Group 4).
While a platform technically allows anyone to see and interact with every user's content, there is still a clear difference between a system requesting users to post in particular subforums and one accumulating posts on a news feed. 
The former can be considered closed group interactions because they happen in clearly delineated spaces, and content does not automatically reach users across a platform~\cite{vanraemdonckTaxonomySocialNetwork2021}.
Such architecture entrusts the site administrators and moderators responsible for sanctioning or rewarding certain behaviours, thus \textit{refusing} open network interaction.
Contrary to that, the affordances of Group 3's artefact \textit{encourage} open network interactions, as content moves freely across the platform. 
Such an architecture focuses on the sender instead of an overarching group space like a subforum.
This puts more responsibility and power onto the senders, as they have more possibilities for intervention and structuring the debate~\cite{vanraemdonckTaxonomySocialNetwork2021}.
This shows a contrast to the other left-leaning Group 2, as they chose the older idea of forums as a solution for their platform architecture. 

Regarding \textbf{user interventions}, we observed various approaches and remarkable differences in the way the groups' artefact afforded users to be involved.
The right-leaning Group 1 considered the moderation rules so important that one of their artefacts was a list of written rules.
While their system imposed a governing style to moderation~\cite{rantasilaOutliningApproachesImprove2022}, they still allowed user interventions through the option to report to content and initiatives for new subforums but did not encourage such action, which shows a more neutral stance of \textit{allowing}.
Instead, the left-leaning Group 3 used self-guided moderation strategies with trigger warnings and warning counts (which could lead to banning from the platform)~\cite{rantasilaOutliningApproachesImprove2022}.
Further, together with the mixed Group 4, they thought of structured content input that promotes user in adding justification for their posts.
Thus, both groups' artefacts \textit{encourage} user intervention.

Finally, it seems that groups including left-leaning participants provided more flexibility for users to present themselves, encouraging or at least not restricting \textbf{flexible profile work}~\cite{devitoPlatformsPeoplePerception2017}.
While all groups required a persistent identity from their users, features such as a changeable username (Group 2) or parody accounts (Group 3)  offered users agency to express themselves more freely. Additionally, Group 3 offered users control over the visibility of their content, \textit{encouraging} users to perform self-presentation by catering to different audiences~\cite{devitoPlatformsPeoplePerception2017}.
In stark contrast, the right-leaning group offered the least opportunities to customise profiles or present freely, which indicates an expectation for the users to present persistent roles on their platform and an affordance that \textit{discourages} flexible profile work.

Our study only had one right-leaning and one mixed composition group due to participant recruitment challenges (see foonote~\ref{footnote:parties_not_workign_with_us}).
However, two left-leaning groups already allow us to highlight diversity within the same end of the political spectrum, demonstrating how various other factors may also impact design work.
For example, with profile work, Group 3 was much more creative in terms of attributes (see Table~\ref{tab:user_profile}).
Further, Group 2 subscribes to the forum approach, which separates interactions on the platforms into pre-defined groups and gives less intervention capabilities to individual users. 

\section{Discussion}

We began our work by highlighting a challenge related to how values are often considered in human--computer interaction without focus on political elites and their political ideologies.
We show that political ideologies should be considered as a relevant set of values in the context of technology design.

Overall, there were some commonalities in the design approaches.
For example, all groups identified the need for strong authentication, linking the users real identity with their online profile.
Similarly, all groups included a picture in the user profile.
We also observed differences across the groups.
Some of these differences align with previous research on political ideologies and their characteristics in non-digital domains.
This highlights the relevance to examine political ideologies as part of the design work.

\subsection{Design by the Left and the Right}

Previous work in political science has various characterisations of the economic left and the right:
while somewhat unclear labels, both citizens and experts identify this continuum and can position themselves into it \citep[see e.g.,][]{COCHRANE2019a}.
Using political party manifestos, scholars such as \citet{COCHRANE2019}, \citet{Jahn2011} and \citet{Daubler2022} have identified common characteristics of political parties based on their ideological leaning:
right-leaning parties are characterised with lesser degree of support to state intervention in the economics or to more expansive welfare programs (see Table~\ref{tab:some_design_ideology}, first row).
As these are from political party manifestos, these do not directly map into the design of digital services.
Rather, they broadly characterise the division between the left and the right in their ways to approach societal questions.

\begin{table}
    \centering
    \begin{tabular}{p{.15\textwidth}p{.4\textwidth}p{.4\textwidth}}
    \toprule
    & The left & The right \\
    \midrule
    Political party manifestos &
    \begin{itemize}[leftmargin=*]
    \item the need for fair treatment of all people, special protection for underprivileged, and end of discrimination;
    \item decreasing military expenditures and disarmament;
    \item preservation of natural resources against selfish interests and other green policy ideals,
    \item support for welfare state;
    \item need to spend money on museums, art galleries, and other cultural and leisure facilities;
    \item favourable reference to underprivileged minorities who are defined neither in economic nor in demographic terms;
    \item support for market regulation and direct government control of economy, as well as state intervention into the economic system.
    \end{itemize}
    & 
    \begin{itemize}[leftmargin=*]
    \item support for free enterprises and trust on market-based dynamics overall as well as personal initiative as well as support for traditional economic orthodoxy and a need for efficiency and economy in government and administration;
    \item support on economic incentives to boost enterprise and entrepreneurship as well as support for free trade;
    \item reduction of state-sponsored social service or social security scheme and limiting spending on education;
    \item support for traditional moral values, such as prohibition, censorship, and suppression of immorality and unseemly behaviour;
    \item less support for labour groups, working class, unemployed;
    \item enforcement or encouragement of cultural integration
    \end{itemize} \\
    \midrule
    Digital design principles &
     \begin{itemize}[leftmargin=*]
         \item Paternalism

         \item Rich identity and profile work
     \end{itemize} &
     \begin{itemize}[leftmargin=*]
         \item Freedom and responsibility
         \item Competitive logic
     \end{itemize}
     \\ 
     \bottomrule
    \end{tabular}
    \caption{Social media design principles for the left--right axis}
    \label{tab:some_design_ideology}
\end{table}

Moving from political manifestos to digital artefacts and design principles in them (see Table~\ref{tab:some_design_ideology}, second row).
\citet{Gron2020} suggested that the left--right dimension appeared as a tension between \emph{paternalism and freedom}.
More extensively, they showed acceptable design approaches seemed to be rooted in the values of each party, for example, communality was preferred by the liberal-conservative and agrarian Centre Party.
Our work continues to examine this area, showing alignment of our work with the work on political manifestos and previous work on digital artefacts and ideologies.

We observed a difference in design choices regarding \textit{popularity-based visibility}that can be interpreted to correspond group's ideological compositions.
The right-leaning group suggested affordances encouraging popularity as a driver for content visibility, which seems to point towards an underlying idea of the platform as a marketplace for ideas competing for attention.
Both left-leaning groups explicitly discouraged such an approach, even criticising such approaches.
This difference continues the well-established distinction of the economic left and right in the competitive logics: left-leaning parties do not in general prefer it, while right-leaning parties believe that open markets and competition leads to positive gains to the society (see Table~\ref{tab:some_design_ideology}, first row).
However, we move this observation from the party manifestos and non-digital choices to the digital service design.

Similarly, we observed differences in the formulation of \emph{user identity}.
Both left-leaning groups appeared to allow and encourage more flexible profile work, while the right-leaning group and mix-membership group limited the options for users to present themselves.
In part, this difference may speak to different conceptualisations of how people form social groups.
For left-leaning parties, group identities have historically been important as attributes which unite people: they were organised as \textit{class-mass parties} and therefore their support has followed social class and related occupational status \citep{Gunther2003}.
Still today, collective identities are important for the left; among other things as an approach to examine and organise around societal fairness \citep{Bernstein2005}.
Therefore, it is reasonable to argue that they may embrace a more nuanced representation of identity.

Overall, we further the argument that political ideals of the left and right guide how digital topics are approached (see Table~\ref{tab:some_design_ideology}).
We see that similar themes touched on non-digital domains, such as party manifestos, can also be identified in the design of digital artefacts and their affordances.
This highlights how political ideologies may -- at least partly -- shape how participants perceive the problem
and choose what the acceptable solution spaces are for such problems.
Our findings on competitive logic and identity and profile work are aligned with the political manifesto work, but also show how profound aspects of social media design are connected with ideologies:
these relate to well-established core affordances of social media services.

\subsection{Beyond Left and Right on the Ideological Spectrum}

As highlighted in Section~\ref{sec:political_ideologies_as_values}, economic left--right is one ideological dimension; but there are other ideologies as well.
Our analysis and outcomes directly relate to our participant selection process, where we sought groups along the economic left--right dimension for this study.
However, following the two-dimensional model, we know that Group 3 stands out as the only one with a pronounced liberal-leaning ideological measurement (see Table~\ref{tab:group_compositions}).
Other groups had only a slight leaning towards the liberal end, but had a high variance within the group of participants, indicating that the groups were more heterogeneous on this ideological dimension.

Among the observations not explained above was Group 3's choice to focus on an algorithmically and profile-based structured social network that allows open network interactions.
The ``forum'' solution by the other groups follows a more traditional approach of structuring the interaction in clearly defined subgroups, which limits network effects.
This might even signify an underlying agreement of slightly more conservative values across party lines and a difference between the two left-leaning groups.

Given our small sample, our evidence is only an indicator and requires further research.
Nonetheless, we make this remark as a response to how the HCI and CSCW research community engages political ideologies with only one dimension \cite[e.g. as in][p. 2607]{zhangModelingIdeologyPredicting2015}.
While this may be acceptable within the United States, the narrow perspective on ideologies cannot be sustained in a global community -- even Western Europe differs from this in the additional value dimension.
In addition, political scientists have highlighted that the ideological landscape is currently changing \citep{fordChangingCleavagePolitics2020,Jahn2023}
, thus requiring sensitivity on this topic.

\subsection{On the method: How to examine ideologies of artefacts?}
\label{sec:method_discussion}

It is often challenging to work out design reasoning from presentations and artefacts; as we note, the framework we used for analysis \citep{Hasinoff2021} was not specific on how to make this move.
Similar challenges also exist for other works which seek to extract ideologies from artefacts \citep{Light2016}.
In response to this challenge, we expanded the analysis approach and embarked on interpretative work on the affordances the participants had included in their designs by reflecting on them using our knowledge of the \textit{self-described ideological leanings} of the designers as a lens for analysis.

Self-described ideological leanings provide us an additional detail for interpretative work; but in qualitative research there are position towards the interpretation  \citep[e.g.,][]{Soden2024,Braun2019}.
\citet{Willig2017} divides interpretation on a continuum between \textit{suspicious} and \textit{emphatic} approaches.
Our approach falls towards the suspicious end of the continuum: the researchers work to connect and interpret observations to wider theoretical perspectives instead of focussing on how participants themselves provide the meanings; this latter is referred to as the emphatic approach.
To ensure transparency in this work, we have used excerpts from the research material to justify our interpretation and used established works in political science to further situate our findings in established body of work.

An emphatic approach would have been possible as well: we could have asked the participants to directly explain their choices from each member's political viewpoint.
In an individual setting, \citet{Gron2020} used such more emphatic approach to elicit an explicit connection with the ideology of the party and the party and the choice of design.
However, as our study was a collaborative design work, we did not want to prime participants at any point of time to focus on their parties' political viewpoints.

While a suspicious approach limits our outcomes (see next section), it allows us to examine how design work is discussed and when and how political alignments step into the discussion and design choices.
Any overt priming to political parties may have lead the design workshop to become more a debate of party positions than to work towards design itself.
At the same time, as the self-described ideological positions were known to researchers, we build a bridge from ideologies to artifacts.


\subsection{Limitations and Future Work}
\label{sec:limitations}

There are several limitations to our study:
most vitally, our sample is small, and there is a lack of right-leaning and mixed groups, inhibiting a more extensive within-ideology comparison.
We re-iterate that this was due to the challenge of reaching out to a particular political party, thus limiting our access to right-leaning participation.
Access is a known problem when studying elites \citep{Jupp2006}. 
Nonetheless, the small sample limits the generalisability of our findings.
Similarly, we acknowledge that some results might be due to the difference in the groups' social networking experiences, age, design experience, or other compounding factors.
For example, it may be that more experienced users of social media could reflect on various existing design approaches and thus have wider approaches in their minds.
Teasing these differences would require a much more extensive number of groups; previous studies on the impacts of political ideology on deliberation had 32 groups and 256 participants \citep{Gronlund2015}.
However, while our sample is clearly underpowered to draw such conclusions, we connect our findings to previous work regarding political ideologies to give them a stronger foundation.
Our results highlight that similar patterns observed in the non-digital domain translate as design inspirations in the digital domain.

Second, political ideologies are highly contextualised.
The outcomes may be different in another country or may change temporarily as political ideologies and their thinking evolve \citep[for summary, see][]{COCHRANE2019a}.
For example, \citet{Nelimarkka2019} show that acceptable polarisation mitigation strategies for social media differ in Finland and the United States.
These two issues call for more extensive studies and comparative approaches to examine how political ideologies shape potential design language, which is beyond the scope of this study design.

Finally, this analysis focused only on the final artefacts and their presentation.
Further research could examine group dynamics during the design stages.
It would be possible to analyse who introduced design ideas, how those were discussed and if those discussions led to their incorporation in the final artefact, focussing on the interaction between participants to understand themes such as power and tensions \citep[previously studied in more generic participatory design by][]{Bratteteig2012}.
We envision this direction as an area for future work to gain a nuanced perspective on how the design ideas emerge and then are shaped by the group into the final design outcomes examined in this work.

\subsection{Implications to the Design of Social Media}

The CSCW community has worked extensively on the design of social media to mitigate societal concerns such as
political polarisation \citep[e.g.,][]{Nelimarkka2018a,Munson2009},
derogatory language use and poor argumentation \citep{Seering2019a,Jia2024,Kriplean2012},
or 
lack of civic control in these systems \citep{Zhang2020,Fan2020}.
Given these concerns, there are proposals on alternative value bases for social media services.
Various works have examined how
algorithmic changes \citep[e.g.,][]{Jia2024,Combs2023,Feltwell2020,Garimella2017}
or alternative design approaches \citep[e.g.][]{Semaan2014,Semaan2015,Nelimarkka2018a} can mitigate these concerns.

However, these approaches have not always gained political support due to ideological conflicts.
The lack of political support is most prominently seen in the debate on moderation and free speech:
\citet{Haapoja2020} elaborates how the deployment of a hate speech detection system produced a counter-movement from those who considered it too limiting for political expression;
this divide aligned with the tensions between political ideologies.
These examples show how the ideological divisions -- similar to those observed in our study and \citet{Gron2020} -- have direct implications on how a service operates.

Therefore, the research community working to propose alternatives should diversify their ideological considerations to capture design space more comprehensively.
For example, the recent proposal to order the news feed based on a \textit{societal objective function}~\citep{Jia2024} instead of the engagement-based news feed omits that specifying such a function might be impossible in a pluralistic society.
Their work chooses the function from political science work on anti-democratic behaviours.
However, working the function out from political sciences does not make it free from values.
For example, conceptualisations of democracy differ widely \citep{held06}, and thus what counts as anti-democratic depends ultimately on a choice about values.

When aiming to `fix' social media's downsides, we must acknowledge that even clearly ideologically motivated work could benefit from examining alternative ideological stances.
Fixing efforts may overlook how technologies manifest values, as choices like free self-presentation or competition for visibility are entangled with established political values, as shown in this research.
One potential direction is to form deliberative processes where a service is developed in a mixed ideology setting vital to ensure the acceptability of the service across the ideological spectrum.
Our mixed composition group already seemed to integrate both more left-leaning and right-leaning principles into their artefact.

Alternatively, we may need to conclude that a service cannot be redesigned to cater for all perspectives. 
Each service has their users and context of use.
For example, users of currently conservative Truth Social may find some design directions unacceptable overall, and might not find it relevant to incorporate more liberal perspectives to the design work at all.
Similarly, cultural differences may impact acceptable and unacceptable design directions \citep{Nelimarkka2019}.
Indeed, recent changes  due to Elon Musk's transformation of Twitter into X and related changes e.g., on moderation have sparked similar migration of users to other services (e.g. BlueSky or Mastodon).
Hence, there is always the avenue to  design new services to provide a space for those who feel not aligned with the platform.
Furthermore, approaches such as federation would allow  platforms share some areas of the content.



\subsection{Implications to Politics and Design}

Beyond the specifics of social media design, our work emphasises the importance of political ideologies as value sets, showing that the design approaches chosen can be interpreted as being rooted in broader ideological aims held by people.
It has been understood that policy and politics are essential parts of design activity \citep[e.g.,][]{Jackson2014,Kannabiran2010} and there are various works which focus on designing from a particular ideological stance \citep[e.g.,][]{Fox2019}.

We expand this discussion by building a connection between the democratically elected political representatives affiliated with political parties and technology design beyond mere regulation.
Traditionally, political action in the technology domain has focused on regulating the relationship between users and platforms, such as through GDPR or the AI Act, and controlling cooperation between platform companies, such as in antitrust cases.
Also here, political ideologies are at the core of these actions.
Political ideologies shape what goes into regulations and laws, and the digital is increasingly an area for politics.
Indeed, political party manifestos clearly differ on how they address the democratic potential of digital platforms, and organisations of labour and capital relationships differ \citep{Guglielmo2024}.
\citet{Guglielmo2024} argues that these create new lines of tensions along which parties differentiate each other. 
Beyond these ideas of how the digital is seen, our work indicates a more general design language that outlines acceptable affordances in technology design per political ideology.


Therefore, our conclusion from the prior section for a need to diversify ideological perspectives in the design and research process applies not just to social media but more generally to design, particularly to areas where societal implications are clear, and more extensively to value-sensitive and critical design approaches.
Current the academic discussion on ideologies has mainly focused on two streams: 
capitalism-critique and feminism \citep[e.g.,][]{Keyes2019,Bardzell2010}
or even see ideologies as 
``constraints'' to be resolved~\citep{burrellIntroductionSpecialIssue2024}.
We call for expanding how we approach and acknowledging ideologies:
ideologies should be explicit, both in the data and author's position. 

This may require asking participants to identify their ideological stance, either directly using dimensions like economic left--right, or through value statements, and reporting these as background factors similar to commonly used age, gender, and race/ethnic background; and ensuring that ideological leanings are transparent.
Additionally should be reflective about how their values are placed in the spectrum of ideologies.
Similarly, author positionality, particularly on various critical design approaches, such engage more deeply with ideological commitments and if possible, reflect how other kinds of commitments may manifest themselves.

Underlining both of these observations, we have used the conceptualisation of political ideologies as a springboard for more extensive engagement with politics and policy.
As our study indicates, this conceptualisation and its relationship with governments and politics may provide a fruitful angle to attack the policy--design gap.

\section{Conclusions: Towards Political HCI?}

Our study explored the design of social media services with four groups formed by politically aligned elites.
By examining the artifacts produced, we observed that some of the design choices were aligned with elite's ideologies.
Certain features were preferred in groups with right-leaning members, while other features were favoured by left-leaning members.
Our findings underscore that political ideologies represent a coherent set of values even in the design domain of digital systems, extending and generalizing existing political science research on ideologies.

Political power increasingly influences the development of information technology and social computing systems through legislation and government funding for technology development.
Beyond this direct influence, political ideologies manifest as broader sets of values, raising the challenge of ensuring more democratic representation of these values during system development.
Therefore, social computing researchers must engage with political science and the study of political ideologies.
This engagement will help understand, unpack, and make visible the political power embedded in these systems, fostering public debate and further examining digital systems' societal impact.

\bibliographystyle{ACM-Reference-Format}
\bibliography{extra,MattiNelimarkka,FelixEpp}

\end{document}